# Comparison of measuring the cloud top height by a radiosonde with an optical sensor and an airplane.

A.V Kochin

*Central Aerological Observatory*

*141707 Russia, Moscow region, 3 Pervomayskaya str., Dolgoprudny*

*Alexander Kochin Email: amarl@mail.ru*

ABSTRACT

Currently, the cloud top height (CTH) is measured by remote satellite and radar methods. Radar methods detect only clouds with precipitation, so their information about the optical cloud top height has little reliability. In addition, any remote method must be monitored. The CTH is most accurately measured by aircraft methods. However, its use for measuring the height of cumulonimbus clouds is almost impossible. The paper presents the results of comparing the measurement of the CTH using a radiosonde with an optical sensor and an aircraft with a special laser locator. The measured heights differed from each other by 30 meters, which corresponds to the accuracy of the equipment used. The purpose of the work is to monitor the quality of satellite and radar methods for measuring the CTH.

1. ## Introduction

The main source of information about clouds is remote sensing, in particular, radar and satellite methods. The existing network of centimeter-range meteorological radars is capable of detecting only clouds with precipitation (Doviak and Zrnich, 1984). Based on decoding optical images at different wavelengths, satellite methods allow detecting clouds and determining the CTH. However, these methods are subject to CTH measurement errors, and sometimes cannot detect clouds (Mokhov 1994). It is especially difficult to determine the presence of clouds in the polar regions. This is due to the fact that the optical characteristics of snow and ice are close, and the CTH in these regions is small (Esau 2015). As a result, the reliability of remote sensing methods decreases. In this case, the availability of an independent contact source for obtaining data is of paramount importance. A radiosonde can be used to deliver a sensor to detect clouds. The worldwide network of the upper-air stations includes about 800 stations around the world that regularly launch radiosondes, so monitoring can be carried out continuously in different parts of the world.

A cloudy environment differs from a cloudless one by the presence of water droplets or ice crystals. This phenomenon changes the optical properties of the medium. The study of optical processes in the atmosphere was carried out using radiosondes (Asano 2004; Kondratiev 1964; Kostyanoy 1975). To ensure high-quality measurements, a system for stabilizing the angle of inclination of the optical radiosonde sensor was used. To study the radiation balance, experiments were also carried out using a tilt angle stabilization system (Philipona 2012). The use of optical sensors based on a conventional photodiode without a tilt angle stabilization system was also proposed in the paper (Nicoll and Harrison, 2012). The proposed system can be considered an optimal solution. A radiosonde is a disposable device, so the main requirement for sensors is their low cost.

**2. Description of the experimental equipment.**

The use of optical sensors of the visible range based on a conventional photodiode without a tilt angle stabilization system was proposed in an paper (Nicoll and Harrison, 2012). However, the obtained data on CTH were not confirmed by an independent measurement. We independently used a similar solution (Kochin 2021), but we were able to confirm the measurement results by comparing them with aircraft data. A conventional commercial photodiode FD-256 (Russia) was installed on the radiosonde. Its operating range was from 0.4 microns to 1.1 microns. The photodiodes were tested in a thermal chamber with the possibility of cooling to a temperature of -70°C. The experiments were carried out using radiosondes manufactured by MODEM M2K2 DC (France). The sampling rate was 1 Hz. Optical visible light sensors were assembled in such a way that the maximum of the radiation pattern was oriented along the guide connecting the radiosonde and the balloon (Fig. 1).

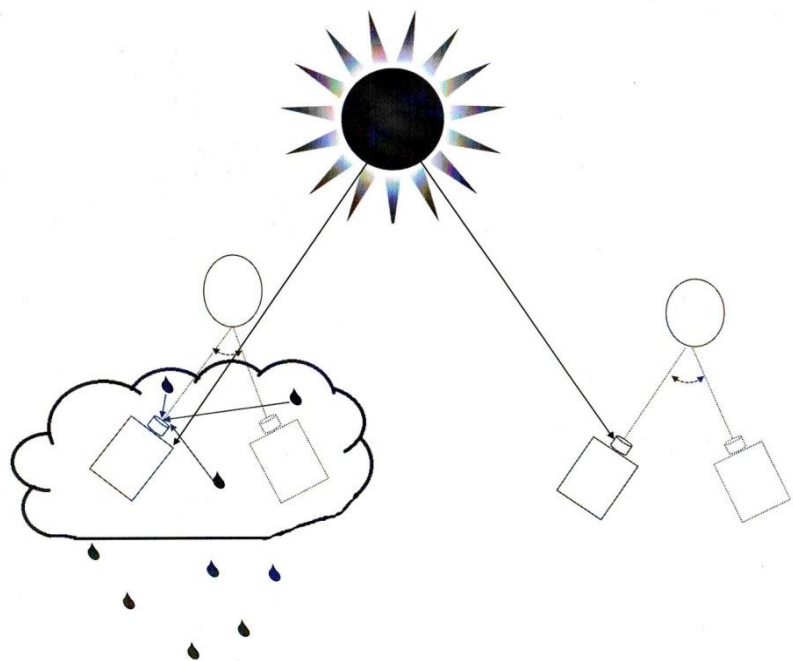

Fig.1. The principle of measuring the CTH.

The change in the intensity of the incident sunlight is weakly dependent on the altitude of the flight. When a radiosonde flies through a cloud, the intensity of direct light decreases, the intensity of light reflected from the cloud particles increases. The change in the nature of the signal fluctuations was used to detect clouds. The radiosonde balloon system swings like a pendulum during flight. The average angle of deviation from the vertical is 18° (Dubovitsky 2015). As in the work (Nick), it was assumed that this would cause noticeable fluctuations in the signal with a clear sky, and in the presence of clouds, the fluctuations would be small. This hypothesis turned out to be not entirely true, as will be shown below.

**3. Observation results**

Radiosondes with optical visible light sensors were launched at the standard time of 12 GMT at the upper air station Dolgoprudnaya (index 27713) near Moscow (55.93 N, 37.52 E). The launch time was scheduled for 15:00 local time. The height of the Sun above the horizon varied from 30 to 50 °. More than 20 launches were carried out in various synoptic conditions (Kochin, 2018). The raw data is shown in Figure 2.

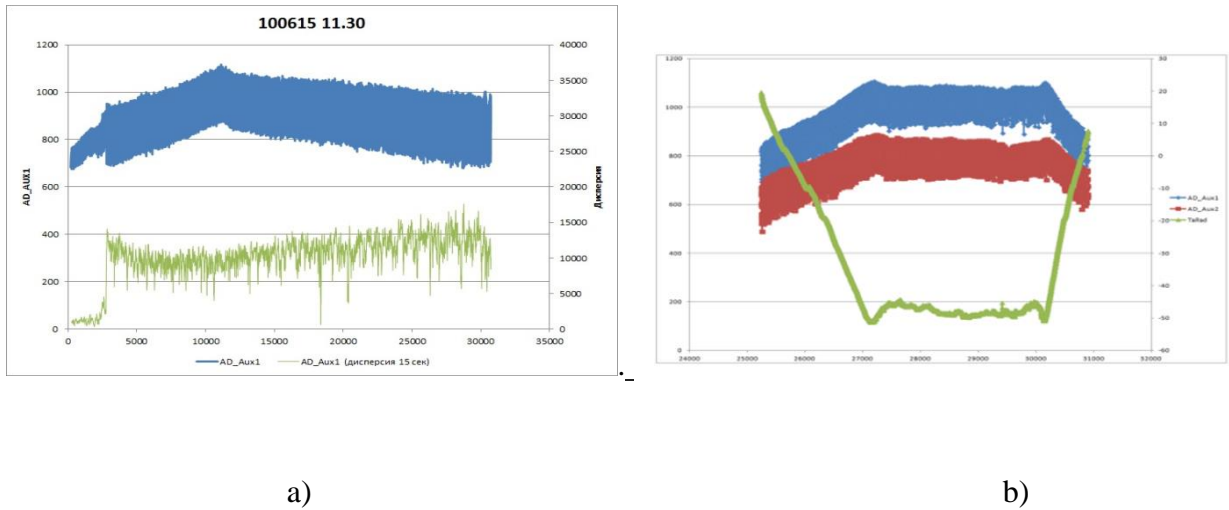

a)                      b)

Fig.2. On the left shows the type of signal in the presence of clouds. The blue color shows the raw signal of the optical sensor of the visible range, and the green color shows the variation of the signal. Horizontally, the height above sea level in meters, vertically, the signal level in ADC code units. On the right, the signal of the optical sensor of the visible range (blue line) and the near-infrared range of the order of 1 microns (red line), the temperature is shown in green. The scale on the right is in degrees Celsius.

The variation value of the optical sensor signal was obtained from an array of 20 values (Kochin 2021). If the length of the array is less than 20 values, the variance fluctuations increase dramatically. 20 values correspond to a spatial resolution of 100 meters, since the radiosonde rises at a speed of 5 m /s, and the sampling frequency is 1 Hz.

In the experiment, the measurement of CTH was carried out using a lidar (laser rangefinder) from the Yak-42D flying laboratory of Roshydromet (Borisov 2012). Figure 2a shows the raw signal in this case. The radiosonde crossed the upper boundary of the cloud cover at the moment when the plane was flying over it. The distance between the aircraft and the radiosonde was about one kilometer. According to the aircraft, continuous layered clouds with a flat upper boundary were observed. The scheme of the experiment is shown in Fig. 3.

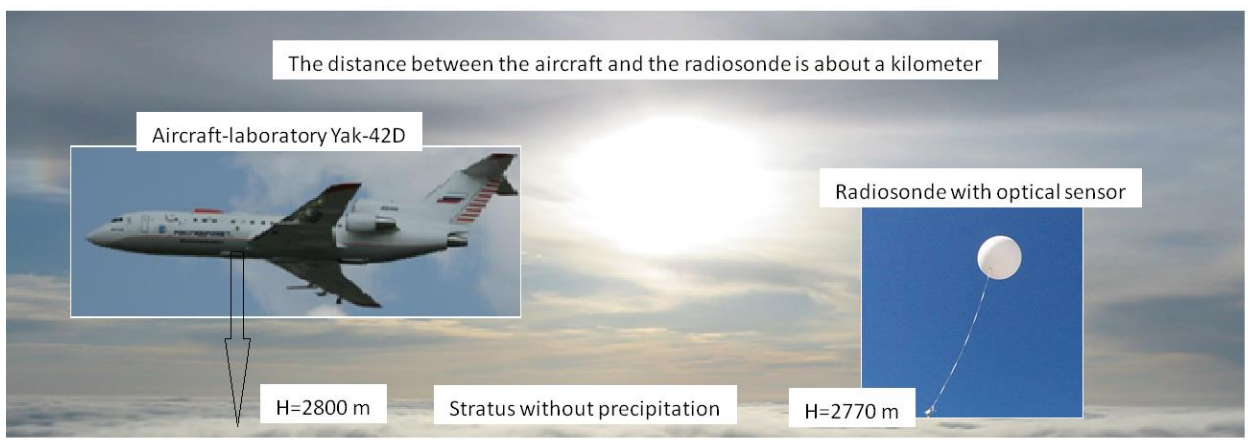

Fig.3. The scheme of the experiment comparing the data of a radiosonde with an optical sensor and an airplane.

The variation of the optical sensor signal increases sharply at a certain height. This is because the radiosonde is leaving the cloud. According to aircraft experiments, the optical thickness of layered clouds is 30 meters (Feigelson, 1981). When the depth of immersion in the cloud is 30 meters, the intensity of direct sunlight decreases by a factor of e (e = 2.71). Consequently, the variation of the intensity of the light incident on the photodetector is reduced by at least two times. Thus, the height at which the variation has halved compared to the variation in a cloudless atmosphere is the optical (visible) height of the cloud top. This threshold was used to calculate the height of the CTH. The presence of clouds and the absence of precipitation were detected both by visual observations and by research aircraft. According to the aircraft, the CTH was 2800 meters. According to the calculation according to the above algorithm, the height of the CTH was equal to 2770 meters. The difference with the aircraft data was 30 meters.

The behavior of the signal in clouds without precipitation has not yet been explained. At a distance of 100 meters or more, direct solar radiation should decrease to almost zero. The variation should also decrease to zero, but this does not happen (Fig.2a). The only explanation is the repeated emission of small particles. Layered clouds consist of cloud particles with a characteristic size of about 5 microns. They do not reflect light back, but scatter it forward (Van de Hulst, 1957). Perhaps this fact leads to self-focusing of direct solar radiation, when cloud particles play the role of re-emitting points. This fact is confirmed by other radiosonde launches and is not an experimental error. An indirect confirmation of the self-focusing effect is the observation of astronomical objects that are visible through layered clouds (Kochin, 2019). If the direct radiation of stars was determined only by attenuation, then observations would be impossible.

The reason for the signal fluctuations in the clear sky also remained unclear. Figure 4 shows the wavelet spectrum of the optical sensor signal.

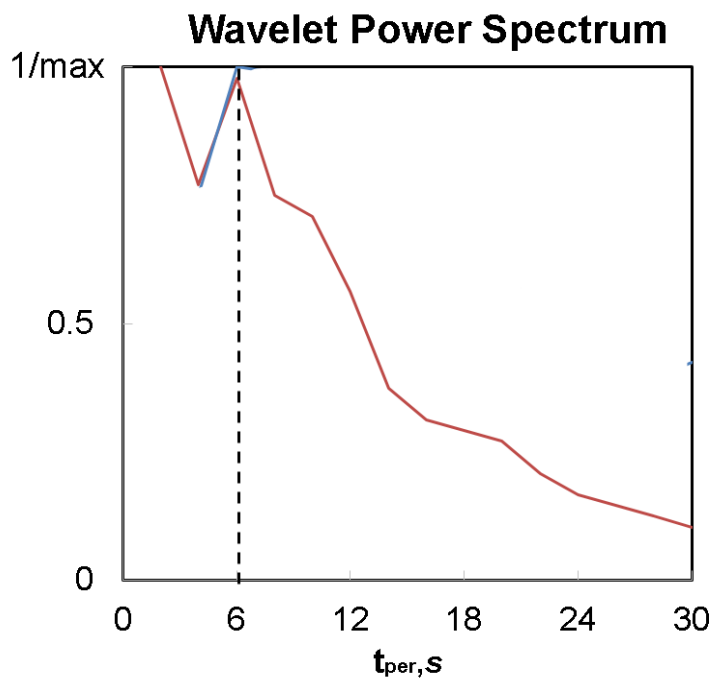

Fig. 4. The wavelet spectrum of the optical sensor signal.

Orientation changes should be displayed as a peak at 6 seconds (the oscillation period of the radiosonde-balloon system). However, the spectrum of the optical sensor is close to the white noise spectrum with a small peak for 6 seconds, which cannot be explained by a change in the orientation of the sensor. Similar results were obtained in (Philippona et al., 2012), which used a sensor with a viewing angle of 180° and a system for stabilizing the angular position of the radiosonde. Changes in the angle of inclination of the radiosonde were absent in these experiments, but similar signal fluctuations were observed in a clear sky.

**4. Conclusions.**

As a result of the measurements carried out, it was found that it is possible to measure the height of CTH with an accuracy of at least 50 meters using an optical sensor during the daytime. This is especially important in polar regions, where the detection of low clouds using satellite methods has low reliability. Ease of use and low cost of consumables are crucial factors for devices used on the meteorological network, so the low cost of the optical sensor will make it suitable for the network. Most modern radiosondes allow you to connect additional sensors and archive the measurement results. The software for calculating variance and determining CTH can be installed remotely, since all stations transmit data over the Internet. Thus, the introduction of a visible light sensor can be carried out without significant additional costs.

**Acknowledgments.**

The author thank the staff of the Dolgoprudny upper-air sounding station for their assistance in conducting radiosonde launches.